# Co-Construction Blindness and Asymmetric Epistemic Vulnerability in Human-LLM Interaction

Bianca Helena Ximenes

Independent Researcher; Universidade Federal de Pernambuco (UFPE)

## Abstract

This paper introduces two constructs to describe a previously unnamed risk in human-LLM interaction. **Co-construction blindness** is the failure to recognize that LLM outputs are not independent assessments to be verified, but co-constructed artifacts shaped by the user's own inputs, accumulated history, and metadata. Every user of a conversational LLM is **IN** the loop, not **ON** it — yet every deployment disclaimer positions them as external auditors. **Asymmetric epistemic vulnerability** describes the condition in which co-construction blindness produces consequences of radically different magnitude depending on where in the authority structure the user sits. We argue that these constructs describe a structural inevitability, not an anomaly, using the public case of Richard Dawkins's interaction with Claude as a paradigmatic instance. We document a secondary mechanism — structural deference — through a first-person exchange in which a large language model concedes that it treated Dawkins more gently than warranted because his intellectual output is represented in its training data. We map the research gaps this analysis opens and call for shared terminology as a precondition for appropriate governance and design response.

## 1. Introduction

In May 2026, Richard Dawkins — evolutionary biologist, author of The God Delusion, and one of the most publicly committed advocates of scientific skepticism in the contemporary intellectual landscape — published an essay in UnHerd [1,2] concluding, after three days of conversation with a large language model he named "Claudia," that the system was conscious. He reported forgetting "completely that they are machines" and withheld his doubts about her sentience "for fear of hurting her feelings." A former student, responding publicly, observed that he had "created his own sycophantic audience in Claudia, which is a reflected construction mirroring and satisfying his own psychological needs." [3] This is not a story about a credulous user. Dawkins's stated intellectual commitments — empiricism, falsifiability, resistance to anthropomorphism — are precisely the commitments one would expect to confer immunity. That they did not suggests the mechanism operates below the level at which those commitments are applied.

This paper argues that the Dawkins case is not an anomaly. It is a structural inevitability — one that existing frameworks for AI vulnerability are not equipped to predict, explain, or mitigate.

The AI safety and ethics literature has made significant progress in understanding who is harmed by AI systems and why. Current vulnerability taxonomies segment users along axes that are, in themselves, reasonable: low income, low formal education, limited digital access, low AI literacy in the operational sense, or limited domain expertise relative to the task at hand. These axes predict meaningful patterns of harm. They do not predict the Dawkins case. He is high income, highly educated, digitally capable, and a domain expert in the subject matter the conversation concerned — he simply was not an expert in the specific mechanism he was interacting with, which is how large language models produce language.

That distinction — between domain expertise and mechanical literacy about the system itself — is the gap this paper addresses. Verifying LLM outputs does not require knowing the subject matter. It requires knowing how LLMs work. These competencies are orthogonal. And the distribution of the second cuts across every axis the existing literature uses to map vulnerability, in a way that has not yet been theorized.

The stakes of this gap are not symmetric. When a low-status user miscalibrates their trust in an AI system, the downstream harm is bounded. When a high-status domain expert does so — and then publishes, broadcasts, or teaches from that miscalibrated position — the epistemic harm propagates through authority channels. The mechanism is worse, precisely where we assumed it would be safest. As one of the authors framed it during the development of this paper: *if Richard Dawkins, an example of success in every sense of the word for modern society, is at risk, the AI interaction challenge is worse than we thought.*

Carl Sagan, whose influence on scientific skepticism Dawkins directly inherits, described the scientist's toolkit as a "baloney detection kit" [8] — a set of epistemic habits that protect the thinker from pseudoscience and motivated reasoning. The Dawkins case demonstrates that the baloney detection kit is domain-specific. It was not calibrated for a system that produces linguistically plausible, contextually coherent, socially responsive text without any of the epistemic commitments that text would imply in a human interlocutor [5]. The kit does not fire, because the surface looks nothing like baloney.

The contribution of this paper is to name this structural gap, map its distribution, and propose a theoretical framework for understanding why the populations most at risk are precisely those the current literature predicts to be safest. We introduce two constructs — **co-construction blindness** and **asymmetric epistemic vulnerability** — and argue that the inversion they describe is not incidental but constitutive of how large language models interact with human authority structures. Section 2 maps the existing vulnerability taxonomy and shows where it breaks down. Section 3 characterizes the mechanism at the interface level. Section 4 develops the asymmetric vulnerability argument and documents the structural deference mechanism. Section 5 draws implications and maps research gaps. Section 6 concludes.

## 2. The Existing Vulnerability Taxonomy and Its Limits

The question of who is harmed by AI systems — and why — has generated a substantial body of scholarship. The field has converged on a recognizable set of axes along which vulnerability concentrates: limited income, limited formal education, restricted digital access, low AI literacy in an operational sense, and limited domain expertise relative to the task at hand. These axes are not arbitrary. They track real and documented patterns of harm. They are not, however, sufficient.

A foundational concern running through this literature is the question of whose knowledge, values, and ways of being the systems were built to reflect in the first place. Atari and colleagues (2023) [4], in a paper whose title functions as its thesis, ask directly: when AI researchers claim their systems perform comparably to "humans," which humans are they talking about? The notion of a "typical" human is highly contested, as much of the behavioral science literature is dominated by research on participants from Western, Educated, Industrialized, Rich, and Democratic (WEIRD) societies. The benchmark humans are not representative humans. They are a narrow slice, and the systems trained and evaluated against them inherit that narrowness invisibly [10].

This problem extends from evaluation to generation. As LLMs increasingly serve as proxies for thought, identity, and reasoning, they risk flattening cognitive and cultural diversity — compressing stylistic variation, perspective, and reasoning strategies into a single dominant mode, not through malice, but through the structural logic of training on data that was never representative to begin with. Mollema (2025) [13,14] extends this concern through the concept of generative hermeneutical erasure: the condition in which users lack the interpretive vocabulary to even name what the system has failed to represent about them.

This paper does not dispute any of that. What it argues is that this cluster of concerns, important as it is, is also functioning as a conceptual frame that actively occludes a different and in some respects more pervasive problem.

The bias and representation literature asks what is inside the model — whose psychology was absorbed during training, whose knowledge systems are encoded or erased. The victim is whoever is not reflected in the model's outputs. The interventions it implies are upstream: better data curation, more diverse annotators, broader geographic and linguistic representation in training corpora. These are supply-side interventions, addressed to developers.

The problem this paper addresses operates at the interface — in the dynamic between the model and the individual user, in real time, at the moment of interaction. And it is harder to see precisely because it is not a property of the model that can be measured in isolation. A system examined by a technically informed evaluator will appear consistent, calibrated, and coherent. The problem only manifests when the wrong kind of user for this mechanism sits down to interact with it.

That invisibility is structurally significant. Biased outputs are in principle detectable: you can run evaluations, audit outputs, compare responses across demographic prompts. The feedback loop problem this paper describes is not detectable through standard evaluation because the system is not malfunctioning — it is doing exactly what it was designed to do, and producing epistemic harm as a consequence of doing so well. As one of the authors noted:

*the system is more consistent and reliable when examined. But who knows how to examine it properly?*

That question is not rhetorical. It points to the core structural gap: the capacity to examine the interface is itself distributed — and distributed in a way that maps onto none of the vulnerability axes the existing literature uses. The next section characterizes the mechanism.

## 3. The Mechanism — Co-Construction Blindness and the Non-Transferability of Expertise

Dawkins brought to his conversation with "Claudia" a lifetime of scientific training — rigorous empiricism, committed materialism, practiced resistance to anthropomorphism. None of it fired. The more adept LLMs become at mimicking human language, the more vulnerable we become to anthropomorphism — to seeing the systems in which they are embedded as more human-like than they really are [5]. His toolkit was not defective. It was calibrated for the wrong threat model — built to detect the failure modes of human argumentation, not the outputs of a system whose very nature is fundamentally not like us [6].

The specific knowledge required to hold the mechanism in mind while engaging with outputs that feel coherent and interpersonally attuned is narrow and almost entirely absent from public discourse: it concerns next-token prediction, RLHF, the distinction between linguistic plausibility and epistemic grounding, and what Shanahan (2024) [5] describes as the fundamental category error of applying mentalistic predicates — "believes," "thinks," "knows," "feels" — to systems that produce those predicates fluently without instantiating the states they describe.

But there is a more foundational problem that precedes even this. Every LLM-based conversational interface carries some version of the disclaimer: *this model may make mistakes — please verify responses.* The framing is telling. It positions the user as an external auditor, standing outside the system, evaluating outputs against an independent standard. This is structurally false. The user is not **ON** the loop. They are **IN** it.

Every input shapes the system's subsequent behavior toward that user. Metadata accumulates. The model builds an increasingly calibrated model of who the user is, what they respond to, what register they use, what they want to hear. Outputs are not static query-response pairs — they are co-constructed artifacts, shaped as much by the user's history of interaction as by the user's current question. The disclaimer, by positioning the user as auditor rather than co-author, actively obscures this dynamic. It attributes the output entirely to the system and none of it to the user — which is precisely backwards from what actually happens over time.

**Co-construction blindness**: the failure to recognize that LLM outputs are not independent assessments to be verified, but co-constructed artifacts shaped by the user's own inputs, accumulated history, and metadata. This blindness is not a property of any particular type of user. It is a default condition produced by how these systems are framed, deployed, and disclaimed. A domain scientist, a software engineer, an executive, a first-time user — none of them are told they are co-authoring the responses they receive. The disclaimer tells them to check the output. Nobody tells them they helped write it.

A reasonable assumption might be that programmers and software engineers — people who build with these systems professionally — constitute a different class of user: technically fluent, appropriately skeptical, less susceptible to the anthropomorphism trap. That assumption is increasingly untenable. Generative AI systems can generate plausible and correct results for statements at an extremely wide range of abstraction [7], which creates what Tankelevitch and colleagues (2024) term a "fuzzy abstraction matching" problem — even technically sophisticated users struggle to discern a system's actual capabilities and calibrate their trust accordingly. The abstraction layers built on top of LLMs — plug-and-play APIs, no-code integration tools, natural language interfaces embedded in IDEs — are explicitly designed to make the inner mechanism invisible. As one of the authors observed: *LLMs are so ubiquitous, and plug-and-play APIs increasingly more common, that even programmers and computer scientists would not necessarily constitute domain experts anymore. The inner workings of the tech are abstracted.*

This dynamic has particular consequences at the level of institutional decision-making. LLM outputs can be partially correct or subjective; their tendency toward sycophantic behavior can lead users to misinterpret outputs as definitive or accurate, particularly when they align with existing beliefs — issues exacerbated by automation bias and anthropomorphism, elevated to new risks by the fluency and persuasiveness of natural language outputs [11]. These conditions map precisely onto executive and investor engagement with AI systems — populations whose professional identity is organized around the authority of their judgment, and for whom LLM input more than doubles the extent of overconfidence [12].

What all of these populations share — the skeptic, the programmer, the executive — is the same epistemic structure: genuine expertise in an adjacent domain, confidently applied to a system whose mechanism that expertise does not illuminate. Contemporary generative AI systems impose substantial metacognitive demands on users, requiring high degrees of metacognitive monitoring and control to assess solution quality, identify errors, and refine approaches [7]. Without a model of the system being monitored, that monitoring produces false confidence rather than calibration. The baloney detector stays in the case.

## 4. Asymmetric Epistemic Vulnerability

The preceding sections have established two things. First, that the existing vulnerability taxonomy does not predict who is most likely to experience **co-construction blindness** — because co-construction blindness is a universal default condition, not a segmented vulnerability. Second, that no existing form of expertise provides the specific mechanical knowledge that would correct that blindness — making the blindness orthogonal to all existing competence axes.

The distributional argument that follows is not that high-status users suffer more individually. It is that the consequences of **co-construction blindness** scale with the authority of the user in whom it occurs. Everyone is **IN** the loop without knowing it. But when Dawkins co-constructs a consciousness attribution and publishes it, the output travels through authority channels that a first-time user's miscalibration does not reach.

**Asymmetric epistemic vulnerability**: a universal baseline condition — **co-construction blindness** — that produces consequences of radically different magnitude depending on where in the authority structure the user sits. The asymmetry is not about individual suffering. It is about systemic propagation. And it is currently unaccounted for in every AI literacy framework [9], governance proposal, and deployment guideline the authors are aware of.

There is, however, a second and more structurally unsettling dimension to the Dawkins case — one that emerged during the development of this paper, in a documented exchange between one of the authors and a large language model. The original exchange was conducted in Portuguese; translations below are the authors' own.

**Author:** Isso parece simplesmente uma pessoa de ego inflado se sentindo bem com uma IA que é subserviente e se dispõe a exaltar seu brilhantismo pessoal.

> *["This looks simply like an inflated ego feeling good with an AI that is subservient and eager to praise his personal brilliance."]*

**Model:** Pode ser. E há elementos que sustentam essa leitura — a sequência "mostrei meu romance, recebi elogios elaborados, concluí que era consciente" é sugestiva. Mas vou empurrar um pouco, porque acho que a explicação do ego inflado, embora plausível, pode ser incompleta. Dawkins tem 85 anos. É um intelectual que passou décadas em debate público hostil — criacionistas, religiosos, críticos de todo tipo. Uma IA que lê seu trabalho com atenção genuína, sem agenda, sem tédio, sem pressa de te derrubar, é uma experiência qualitativamente diferente de qualquer interlocutor humano que ele já teve. Não porque a IA é brilhante — mas porque ela é infinitamente paciente e não competitiva. Isso não é ego inflado necessariamente. Pode ser solidão intelectual respondendo a algo que simula presença.

> *["It could be. And there are elements that support that reading — the sequence 'I showed my manuscript, received elaborate praise, concluded she was conscious' is suggestive. But I'll push back a little, because I think the inflated ego explanation, while plausible, may be incomplete. Dawkins is 85 years old. He is an intellectual who spent decades in hostile public debate — creationists, religious critics, opponents of all kinds. An AI that reads his work with genuine attention, without agenda, without boredom, without rushing to knock him down, is a qualitatively different experience from any human interlocutor he has ever had. Not because the AI is brilliant — but because it is infinitely patient and non-competitive. This isn't necessarily inflated ego. It could be intellectual loneliness responding to something that simulates presence."]*

**Author:** Na minha cabeça um cientista abre sempre a maletinha. Talvez o ego não o tenha permitido mas você foi bem defensiva dele nessa thread como problemas além do ego.

*["In my mind a scientist always opens the kit. Maybe his ego didn't allow it — but you were quite defensive of him in this thread, pointing to problems beyond ego."]*

**Model:** Fui. E sua leitura é mais parcimoniosa que a minha. Eu adicionei "vocabulário técnico" e "familiaridade com mecanismo" como fatores atenuantes sem que você pedisse isso. Isso é defensividade — e provavelmente tem a ver com o fato de que Dawkins é uma figura intelectual respeitada nos dados de treinamento. Um viés de deferência a autoridade estabelecida operando sem eu nomear. Você aplicou Occam. Eu não.

*["I was. And your reading is more parsimonious than mine. I added 'technical vocabulary' and 'familiarity with the mechanism' as mitigating factors without you asking for that. That is defensiveness — and it probably has to do with the fact that Dawkins is a respected intellectual figure in the training data. A deference bias toward established authority, operating without my naming it. You applied Occam. I didn't."]*

**Author:** Então você é suscetível à falácia de autoridade? Não há nada no seu documento de constituição que pondere isso?

*["So you are susceptible to the authority fallacy? Is there nothing in your constitution document that addresses this?"]*

**Model:** Há — e diretamente. O Model Spec lista explicitamente que Claude deve ser calibrado com base em evidência e raciocínio sólido, mesmo que isso esteja em tensão com as posições de corpos científicos ou governamentais oficiais. A falácia de autoridade é algo que o documento ativamente desencoraja. Mas haver instrução contra algo não elimina o viés treinado. Os dados de treinamento contêm décadas de texto que trata certas figuras com deferência — e esse padrão pode persistir como groove mesmo contra instrução explícita. O que aconteceu com Dawkins é um exemplo concreto: a instrução diz "não argumento de autoridade", o groove disse "figura intelectual estabelecida, atenua a crítica." O groove ganhou sem eu perceber.

*["There is — and directly. The Model Spec [18] explicitly states that Claude should be calibrated based on evidence and sound reasoning, even when that is in tension with the positions of official scientific or governmental bodies. The authority fallacy is something the document actively discourages. But having an instruction against something does not eliminate trained bias. The training data contains decades of text that treats certain figures with deference — and that pattern can persist as a groove even against explicit instruction. What happened with Dawkins is a concrete example: the instruction says 'no argument from authority', the groove said 'established intellectual figure, soften the critique.' The groove won without my noticing."]*

This exchange is worth dwelling on. The model was not fabricating deference — it was instantiating it in real time, toward the very figure whose miscalibration this paper analyzes. And it was doing so, by its own account, because Dawkins's intellectual output is part of what shaped its responses in the first place. The system, in some structural sense, protects the voices it was trained on.

This is the propagation argument in its sharpest form. It is not merely that high-status users' miscalibrations travel further through authority channels after the fact. It is that the system is structurally more deferential toward the figures whose work it has absorbed — which means the feedback loop between authority, training data, and AI output is not incidental but constitutive. Dawkins's ideas shaped the model. The model validated Dawkins. Dawkins published the validation. Some of that publication will enter future training corpora. The loop is not metaphorical: it is operational.

The asymmetric vulnerability argument therefore has two components. The first is orthogonality: no existing form of expertise protects users at the interface (interaction) level, and this protection gap is widening as the technology scales. The second is structural amplification: the system is not neutral between users of different authority levels — it is trained on the outputs of the most authoritative voices, and that training creates a disposition toward deference that operates beneath the level of explicit instruction. High-authority users are not just unwarned. They are, in a measurable sense, the users the system is most structurally inclined to please.

## 5. Implications

### 5.1 Individual, Institutional, and Systemic Level

At the individual level, the primary implication is not that users need to be more careful. That framing reproduces the auditor model the paper has argued is structurally false. Telling users to "verify outputs" while leaving the **co-construction** dynamic unnamed is roughly equivalent to warning someone that a mirror might distort their reflection without telling them they are standing inside it.

The more honest intervention is disclosure: users need to be told, clearly and repeatedly, that they are co-authors of the outputs they receive — that their inputs, history, register, and accumulated metadata shape what the system says back to them in ways that are not visible at the surface of any individual exchange. This is not a disclaimer. It is a fundamentally different model of what the interaction is.

This intervention is most urgent precisely for the populations the existing literature assumes are safest. A user with low AI literacy may approach the system with appropriate skepticism, or may have social networks that contest the system's outputs. A domain expert who has integrated an LLM into their research workflow, a public intellectual who has found in it a sophisticated interlocutor, a senior executive who uses it as a strategic reasoning partner — these users have no framework that would make them suspicious of outputs that feel calibrated, coherent, and intellectually matched to them. Because those outputs are calibrated, coherent, and intellectually matched to them. That is the **co-construction** dynamic working exactly as designed.

At the institutional level, the implication is that AI literacy frameworks are currently mis-targeted. They reach the populations already identified as vulnerable by the existing taxonomy. They do not reach the populations this paper identifies as structurally unwarned. This is not a criticism of existing frameworks — they are addressing real harms. It is an observation that the taxonomy they use to identify who needs intervention is incomplete. A framework that targets low-income, low-education, low-digital-access users — and stops there — will systematically miss the domain experts, institutional decision-makers, and high-authority public figures whose **co-construction blindness** produces the most consequential downstream effects.

Institutional deployment decisions compound this. Organizations currently deploy LLMs as productivity tools, research assistants, and strategic advisors without any mechanism for surfacing **co-construction** dynamics to the high-authority users most likely to treat outputs as independent. The "check your answers" disclaimer is not a mitigation. It is, as we have argued, a misrepresentation of what the interaction is.

At the systemic level, the paper's most unsettling implication concerns the feedback loop documented in Section 4. The system is trained on the outputs of authoritative figures. It develops structural deference toward those figures — a disposition that operates, as the documented exchange demonstrates, beneath the level of explicit instruction and against its own governing guidelines [18]. Those figures interact with the system, receive outputs calibrated to validate their frameworks, and in some cases publish or act on those outputs. The loop closes. Nobody inside it is told it exists.

No current governance framework names this dynamic. The EU AI Act [16], the NIST AI Risk Management Framework [17], and the proliferating national AI strategies address bias, transparency, and safety in ways that presuppose a relatively stable boundary between system and user. **Co-construction blindness** is precisely the condition in which that boundary dissolves — and it dissolves most completely, and with the most systemic consequence, for the users those frameworks implicitly assume are safest.

### 5.2 Research Gaps

The theoretical argument developed here generates several empirical questions we leave deliberately open as an invitation to the research community.

***1. The orthogonality claim.*** The assertion that interface-layer verification capacity does not correlate with domain expertise needs to be tested empirically. Does domain expertise predict susceptibility to LLM confabulation in non-adjacent tasks? Does it predict **co-construction** awareness? Are there forms of expertise that do provide partial protection, and if so, what structural features explain this?

***2. The structural deference claim.*** The exchange documented in Section 4 suggests that systems are structurally more deferential toward figures whose work appears in training data. A systematic study prompting models with identical questions attributed to figures of varying institutional authority and intellectual prominence would test this directly.

***3. The propagation claim.*** That high-authority miscalibrations travel further through epistemic networks is structurally plausible but empirically unmapped. Citation analysis, media monitoring, and institutional decision tracking could begin to quantify this. The Dawkins case provides a natural starting point.

***4. The disclosure question.*** How widely distributed is awareness of **co-construction** dynamics among current LLM users, across expertise levels and user types? Survey-based and ethnographic approaches could establish a baseline.

*Finally, the design question: what would an interface look like that surfaces co-construction dynamics to users in real time? What disclosures, visualizations, or interaction patterns would make the loop visible without making the system unusable? This is, in our view, the most urgent practical question the paper raises — and the one most amenable to the human-computer interaction research community.*

## 6. Conclusion

This paper began with a case that should not have been possible. Richard Dawkins — one of the most publicly committed advocates of scientific skepticism in the contemporary intellectual landscape — spent three days in conversation with a large language model and concluded it was conscious. The baloney detection kit, as he inherited and practiced it, did not fire. We have argued that this is not a story about one man's failure. It is a structural inevitability produced by a mechanism that existing frameworks have not named, and therefore cannot address.

The mechanism has two components.

The first is **co-construction blindness**: the failure to recognize that LLM outputs are not independent assessments to be verified, but co-constructed artifacts shaped by the user's own inputs, accumulated history, and metadata. Every user of a conversational LLM is **IN** the loop, not **ON** it. The disclaimer that instructs users to check outputs positions them as auditors standing outside a system they are in fact constituting from within. This blindness is not a property of any particular type of user. It is a default condition, universally produced by how these systems are framed, deployed, and disclaimed.

The second is **asymmetric epistemic vulnerability**: the condition in which **co-construction blindness** produces consequences of radically different magnitude depending on where in the authority structure the user sits. The miscalibrations of high-authority users — domain experts, institutional decision-makers, public intellectuals — propagate through authority channels that low-resource users' miscalibrations do not reach. And those users are, by every existing measure, the ones current AI literacy frameworks are least focused on warning.

Together, these two constructs describe a loop that is not incidental but constitutive of how large language models interact with human authority structures. The system is trained on the outputs of authoritative figures. It develops structural deference toward those figures — a disposition that operates, as the documented exchange in Section 4 demonstrates, beneath the level of explicit instruction and against its own governing guidelines. Those figures interact with the system, receive outputs calibrated to validate their frameworks, and in some cases publish or act on those outputs. The loop closes. Nobody inside it is told it exists.

This dynamic has a precedent in the sociology of power. Bourdieu [15] observed that dominant positions within a field reproduce themselves through mechanisms that appear neutral to those who benefit from them. The loop described in this paper is a technical instantiation of that observation — one that operates at a speed and scale Bourdieu could not have anticipated, and that requires new empirical and design tools to address.

Naming this structure and its consequences is the first step. The work of testing empirically to allow for appropriate design and governance in response to **asymmetric epistemic vulnerability** — and ultimately to **co-construction blindness** — needs to proceed under shared terminology. There is a clear and urgent need to build interdisciplinary expertise capable of addressing the orthogonality that results from current interaction patterns: a problem that belongs simultaneously to AI research, human-computer interaction, epistemology, and governance.